\documentclass[11pt,twoside]{article}

\usepackage{asp2006}
\usepackage{fancyhdr,graphicx,caption,rotating,multirow,lscape}
\DeclareGraphicsExtensions{.eps,.eps.gz,.ps,.jpg,.tiff}

\begin{document}

\title{Searching for Planetary Transits in Globular Clusters - 47 Tucanae and $\omega$ Centauri.}  



\author{David T.F Weldrake}
\affil{Max-Planck Instit\"ut f\"ur Astronomie, K\"onigstuhl 17, Heidelberg, 69117 Germany}
\author{Penny D Sackett}  
\affil{Research School of Astronomy and Astrophysics, Mount Stromlo Observatory, Cotter Road, Weston Creek, ACT 2611 Australia}    
\author{Terry J Bridges}
\affil{Department of Physics, Queen's University, Kingston, Ontario. K7L 3N6 Canada}

\begin{abstract}
We have performed a large ground-based search for transiting Hot Jupiter planets in the outer regions of the globular clusters 47 Tucanae and $\omega$ Centauri. The aim was to help understand the role that environmental effects play on Hot Jupiter formation and survivability in globular clusters. Using the ANU 1m telescope and a $52' \times 52'$ field, a total of 54,000 solar-type stars were searched for transits in both clusters with fully tested transit-finding algorithms. Detailed Monte Carlo simulations were performed to model the datasets and calculate the expected planet yields. Seven planets were expected in 47 Tuc, and five in $\omega$ Cen. Despite a detailed search, no planet-like candidates were identified in either cluster. Combined with previous theoretical studies of planet survivability, and the HST null result in the core of 47 Tuc, the lack of detections in the uncrowded outer regions of both clusters indicates that stellar metallicity is the dominant factor inhibiting Hot Jupiter formation in the cluster environment.
\end{abstract}

\section{Project and Dataset Overview}
Globular clusters are excellent laboratories in which to perform a dedicated search for planetary transits. In a single field of view the brightest globulars present tens of thousands of solar-like main sequence stars, which can be simultaneously analyzed for the photometric signature of an orbiting planet. These clusters allow study into how environment affects Hot Jupiter planet formation and survivability, by placing strong observational constraints on the frequency of Hot Jupiter planets in these non-local environments. Such constraints on short period planet frequencies at low metallicities are not well known, due to the metallicity bias of the nearby bright stars studied by ongoing radial velocity searches. In the case of globular clusters, are the frequencies different, and if so, is stellar density or metallicity more important in determining the frequencies?

To address these issues, we have sampled the outer halos of the two brightest globular clusters in the sky, 47 Tuc and $\omega$ Cen. This work comprises the first dedicated transit survey in $\omega$ Cen, and the first wide-field ground-based survey of 47 Tuc, following on from the HST null result in the cluster core by \citet{G2000}. For 47 Tuc, our results have previously been published \citep{W2005}, placing a strong constraint on the frequency of Hot Jupiter planets (R$\sim$1.3R$_{\rm{Jup}}$) out to an orbital period of 16d.

For $\omega$ Cen, our observations placed constraints on planets with larger radius limits (R$=$1.3$-$1.6R$_{\rm{Jup}}$) with an upper orbital period limit of 7d. Major side-results include the detection of variable stars, with 100 found in 47 Tuc (69 new discoveries; \citealt{W2004}) and 187 found in $\omega$ Cen (81 new discoveries; \citealt{W2006}). In total both surveys present V and I photometry (with astrometry) for a quarter of a million stars. 

Our datasets were produced with the ANU 1m telescope located at Siding Spring Observatory and the 52$'\times$52$'$ field of view of the ANU Wide Field Imager. They are most sensitive to the cluster outer halos where photometric accuracy is higher due to reduced crowding. These parts of the cluster have distinct advantages over the cores; they have small stellar densities and are dominated by low mass stars. We have 56 nights of data covering both clusters, comprising 2100 images, and have photometry for $\sim$53,000 cluster main sequence stars (17.2$\le$V$\le$19.5) with which to search for transits. Our total light curve database was produced via differential photometry and comprises 220,000 stars (14.0$\le$V$\le$22.5). Our photometric accuracy is 2$\%$ at V$=$18.2, and 4$\%$ at V$=$19.0 across the field, after an application of the \citet{T2005} systematics removal package. The whole field except the inner 6$'$ of the core was sampled in both clusters. Considering the detection statistics, such a large photometric database presents an excellent opportunity to detect Hot Jupiter planets and statistically study their relative frequencies compared to the solar neighborhood.

\section{Transit Expectations}
\begin{figure}[!t]
\centering
\includegraphics[angle=0,width=6.4cm]{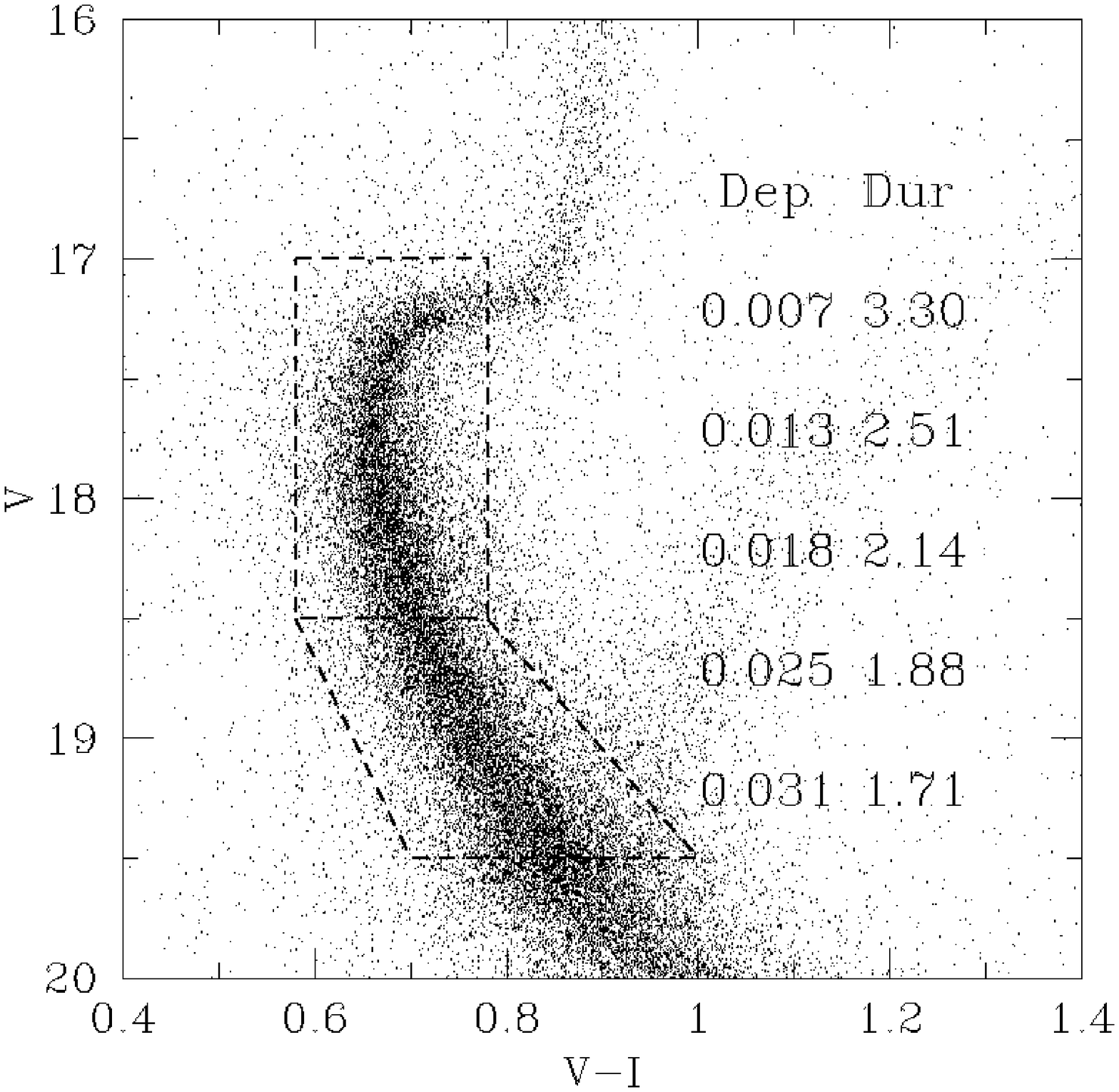}
\hspace{0.1cm}
\includegraphics[angle=0,width=6.4cm]{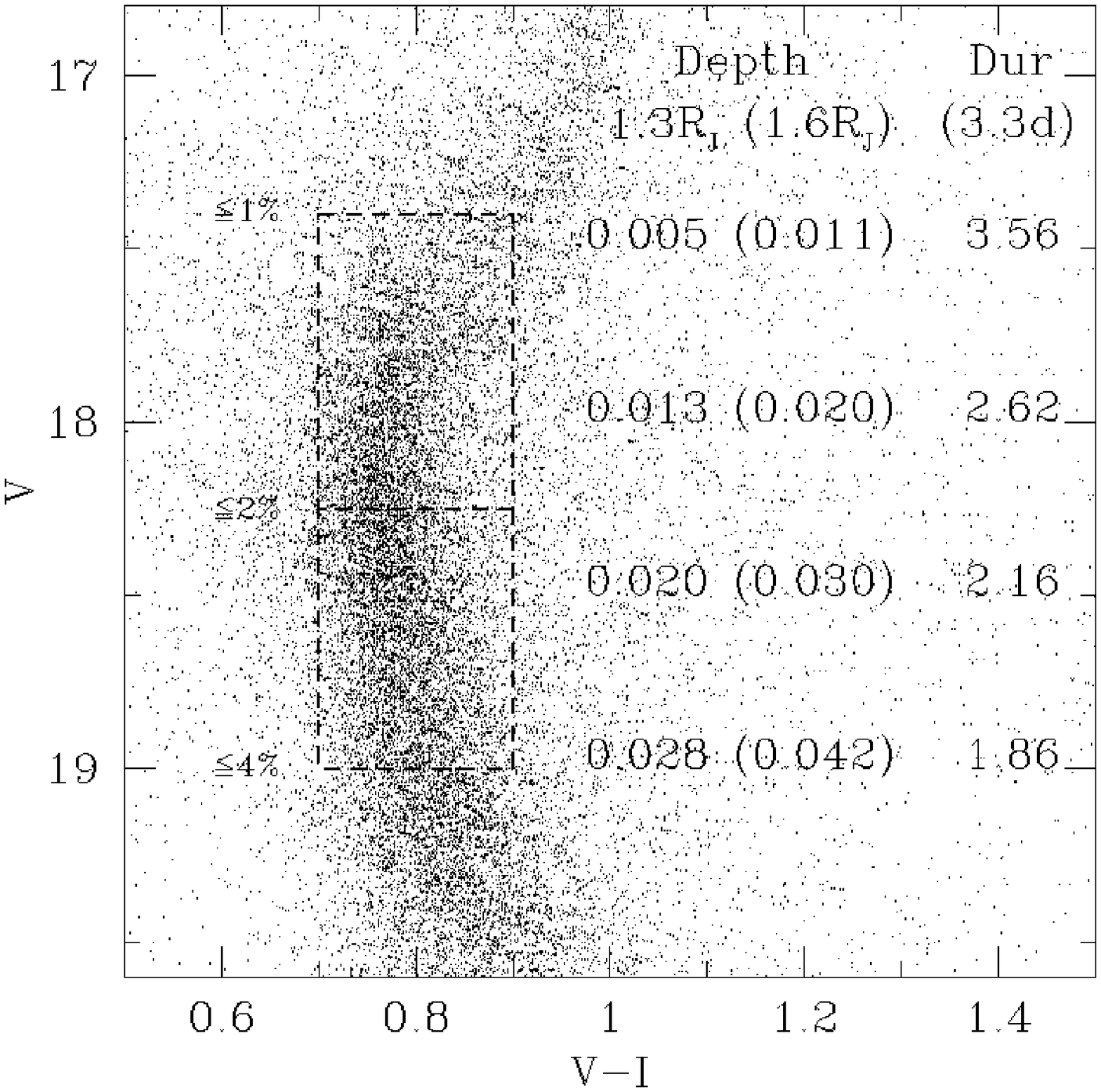}
\caption{The expected depths (magnitudes) and durations (hours) of planetary transits as a function of V magnitude on the main sequence of 47 Tuc (left) and $\omega$ Cen (right) using stellar radius parameters found from \citet{Y2003} isochrones. The duration values assume a fixed orbital period of 3.3d, the most typical period of Hot Jupiter planets in the solar neighborhood. The dotted boxes limit the search range, with the bright limit at the location of 1$\%$ photometry and the lower limit where our uncertainty reaches 4$\%$. With these parameters, simulated transits were produced to determine the statistical transit recoverability in each dataset. \label{expdepths}}
\end{figure}

A significant amount of statistical analysis was carried out on both datasets independently, to determine the transit recoverability and expected planet yields. The first step involves finding the expected depth and duration parameters of Hot Jupiter transits as a function of V magnitude on the main sequence of each cluster. As the cluster stars can all be assumed to lie at the same distance, this was performed by producing \citet{Y2003} theoretical isochrones to simulate the stellar population of each cluster. The stellar radius was found as a function of V magnitude along the cluster main sequence and the resulting transit depth and duration determined for a hypothetical Hot Jupiter planet orbiting with an assumed period of 3.3d (the most typical period of planets in the solar neighborhood) and radii of 1.3R$_{\rm{Jup}}$ and 1.6R$_{\rm{Jup}}$. The results can be seen in Fig.\space\ref{expdepths} for both 47 Tuc (left) and $\omega$ Cen (right). 

The bright magnitude limit for the transit search lies at the location of the cluster main sequence turnoff, as stars brighter than this correspond to foreground stars and cluster stars of rapidly increasing radius, with increasingly undetectable transit depths. For our clusters this limit is V$=$17.2 for 47 Tuc and V$=$17.5 for the slightly more distant $\omega$ Cen. Both clusters display comparable photometric results, and the faint limit to the search is defined as the magnitude at which the photometric uncertainty becomes larger than the expected transit depths. Transit recoverability becomes increasingly truncated after this point. This occurs at V$=$19.5 for both clusters. These limits allow us to sample at least the upper 2 magnitudes of the cluster main sequence for transit signatures. Appropriate transit models can be made to match the radii of this subset of the cluster population, testing the transit detection algorithms and allowing a determination of the expected planet yields.

\section{Monte Carlo Simulations and Transit Recoverability}
A full statistical analysis was performed via extensive Monte Carlo simulations to determine the transit recoverability in our two datasets. Simulated transits were added to actual dataset light curves using the expected transit depth and duration values for each cluster. By using real lightcurves, our simulated transits contain the same photometric information as our real data; this allows us to have a better idea of the effects of systematics on transit detectability, and the effect of the dataset window function on transit visibility.

For our Monte Carlo simulations, we consider that bins of differing photometric uncertainty correspond to stars within a certain magnitude range, and study each in turn. First, we consider that all stars with photometric precision $\le$1$\%$ correspond to brighter stars lying purely in the foreground of the cluster, as cluster stars of this brightness correspond to giants. These foreground stars have sufficiently good photometry to permit transit detection of orbiting companions down to a radius of 0.7R$_{\rm{Jup}}$. The upper limit was arbitrarily chosen to be 1.6R$_{\rm{Jup}}$. Hence, for these brightest stars, simulated transits were added to dataset light curves corresponding to orbiting companions of radius 0.7, 1.0, 1.3 and 1.6R$_{\rm{Jup}}$, with duration corresponding to the period range of 1$-$16d (for 47 Tuc) and 1$-$7d for $\omega$ Cen.
   
The cluster was simulated in two photometric precision bins, one with 1$-$2$\%$ photometry corresponding to the cluster upper main sequence, and one with 2$-$4$\%$ corresponding to the lower main sequence, extending down to our photometric limit. As the photometry becomes worse, smaller planets are no longer detectable, hence for the upper main sequence bin, transits corresponding to orbiting companions of 1.0, 1.3 and 1.6R$_{\rm{Jup}}$ were produced, and only 1.3R$_{\rm{Jup}}$ and 1.6R$_{\rm{Jup}}$ companions for the lower main sequence. To determine sufficiently the transit recoverability, over a million simulated transits were produced with appropriate parameters for the transit detection algorithms to recover.

\begin{figure}[!t]
\centering
\includegraphics[angle=0,width=6cm]{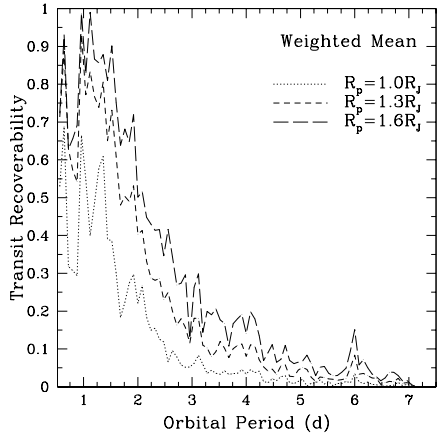}
\caption{The weighted mean statistical transit recoverability for the $\omega$ Cen dataset. This takes into account the individual recoverabilities for differing planetary radii (over-plotted) as well as the number of stars as a function of magnitude along the main sequence. Our data displays good recoverability to an orbital period of $\sim$3d for the larger radius companions.\label{transrec}}
\end{figure}

We used two detection algorithms, the BLS method of \citet{K2002} and the MFA algorithm of \citet{WS2005}. Both algorithms produced very similar recoverability rates, and the MFA algorithm was used as a first-pass test to eliminate many false detections before BLS was used to find accurate period information for the resulting candidates. Any candidates which matched the true period of the input transit, or corresponded to an integer alias were classed as recovered. The recoverability of each photometry bin was determined, and then the weighted mean transit recoverability was found taking into account the relative numbers of stars that lie in each of the bins. 

The resulting transit recoverabilities for the case of $\omega$ Cen can be seen in Fig.\space\ref{transrec}. The recoverability for 47 Tuc has previously been published in \citet{WS2005}. The recoverability for $\omega$ Cen is good in this case out to an orbital period of $\sim$3d, after which it becomes increasingly truncated due to the temporal properties of our dataset. The larger radius companions have higher recoverability, as expected due to their increased transit signal. With this information it is possible to determine an upper limit to the expected number of planets we should detect in our two datasets.

\section{Expected Number of Planets}
For any transit search, accurate calculations must be made to determine the expected planetary yield, in order to place meaningful statistics on the relative frequency of planets in the target field. For stellar clusters, this task is simplified somewhat, as all the target stars can be assumed to lie at the same distance. Theoretical isochrones can give useful information on the physical properties of the stars inside the search regime. Several factors must be considered to find the planet yield, namely the total number of stars, the occurrence frequency of detectable planets, the transit probabilities incorporating the radii of the target stars, and the weighted mean transit recoverabilities - which in our case incorporates residual dataset systematics and the dataset window function.

We determine separately our sensitivity to 1.3R$_{\rm{Jup}}$ and 1.6R$_{\rm{Jup}}$ planets and assume their orbital period distribution to be the same in the clusters as it is in the solar neighborhood. In total, we have data for 22,000 main sequence stars for 47 Tuc and 31,000 for $\omega$ Cen. Our search encompasses the orbital period regimes of both Hot Jupiter and Very Hot Jupiter planets, and thus each type of planet is theoretically detectable. \citet{G2006} present a statistical analysis of the occurrence frequencies for both of these types of planet, calculated from results of radial velocity surveys and transit surveys combined. They conclude that Hot Jupiters and Very Hot Jupiters have occurrence frequencies of 1/310 and 1/690 respectively. These numbers can be used to find the total possible number of such planets that can be associated with the stars in our sample, assuming these frequencies hold for our clusters.

However, not all of these planets will transit, and the transit probability was determined taking into account the radii of the target stars, and the weighted mean radius for the stars in 47 Tuc is 0.94R$_{\odot}$, with 0.96R$_{\odot}$ for $\omega$ Cen. The transit probability was then found for orbital periods from 1$-$16d for 47 Tuc and 1$-$7d for $\omega$ Cen with this stellar radius in bins of 0.5d width. 

By incorporating the previously determined transit recoverability for these same bins, the total number of detectable planets was found by multiplying the factors above and summing over all bins. For 47 Tuc, we expect to find 7 planets (3.3$\sigma$ null significance) of radius 1.3R$_{\rm{Jup}}$ with a period range of 1$-$16d. For $\omega$ Cen, we expect 5.3 planets (2.8$\sigma$ null significance) in the radius range 1.3R$_{\rm{Jup}}$$-$1.6R$_{\rm{Jup}}$ with period 1$-$7d. These are firm upper limits to the expected number. When taken together, these numbers provide a high significance result if no planets are found. A total of 0.7 planets are expected among the foreground Galactic disk stars.

\section{Search Results and Conclusions}
A vigorous search was made of all 53,000 stars with both the BLS and MFA transit detection algorithms. Despite a detailed examination of all candidate detections, we found no planetary transits around any sampled cluster star. This null result has a significance of 3.3$\sigma$ from Poisson statistics for 47 Tuc and 2.8$\sigma$ for $\omega$ Cen, and agrees with the null result of \citep{G2000} in the core of 47 Tuc, indicating that the occurrence frequency of Hot Jupiter planets is reduced in 47 Tuc and $\omega$ Cen. 

Our expected planet numbers assume that stellar metallicity does not affect short period planetary frequencies, contrary to the apparent trend in the solar neighbourhood. By accounting for the low metallicities of the clusters using this trend, the expected number of planets we should find reduces essentially to zero. The dynamics of the clusters are not sufficiently violent to disrupt planetary systems, with a characteristic lifetime of close-in planets estimated to be greater than the Hubble time \citep{Freg2006}. Considering this, our null result suggests that stellar metallicity, not dynamics, is the dominant effect limiting the frequency of short period massive planets in globular clusters, and places an observational constraint on planetary frequency at such a low metallicity. Perhaps the low metallicity does not affect planet formation, but does affect planetary migration. If true, then long period planets should still exist in these clusters, undetectable in our work.

\section{Lupus Control Field}
A control dataset in the Lupus Galactic Plane field was observed for 53 nights in 2005 and 2006 in order to test the observing strategies and analysis methods we performed on our globular cluster datasets, and compare the results. Analysis of $\sim$26,000 stars with photometry better than 2$\%$ is ongoing, and has led to the identification of transit candidates, one of which has excellent prospects for being a new Hot Jupiter planet \citep{W2007}. As no such candidates were seen in either of our clusters, with the same methods employed, these results strengthen the conclusion that short period planets are far rarer in globular clusters than in the general Galactic field.



\end{document}